\documentclass[final,1p,times]{elsarticle}

\usepackage{amssymb}

\journal{Applied Mathematics and Computation}

\begin{document}

\begin{frontmatter}


\author[label1,label2]{Hu Xiong\corref{cor1}\fnref{}}
\ead{xionghu@pku.edu.cn}
\cortext[cor1]{Corresponding author}

\title{Privacy-Preserving English Auction Protocol with Round Efficiency}

\author[label1]{Zhong Chen}
\address[label1]{School of Electronics Engineering and Computer
Science, Peking University, Beijing, P.R.China}
\address[label2]{School of Computer Science and Engineering,
University of Electronic Science and Technology of China, P.R.China}



\begin{abstract}
A privacy-preserving English auction protocol with round efficiency
based on a modified ring signature has been proposed in this paper.
The proposed protocol has three appealing characteristic: First, it
offers \textit{conditional privacy-preservation}: on the one hand,
the bidder is anonymous to the public, on the other hand, only the
collaboration of auctioneer and registration manager can reveal the
true identity of a malicious bidder. Second, it \textit{does not
require to maintain a black list} which records the evicted
malicious bidders. Finally, it is \textit{efficient}: it saves the
communication round complexity comparing with previously proposed
solutions.
\end{abstract}


\end{frontmatter}


\section{Introduction}
\label{sec1}

Electronic auctions are a very popular trading method for
determining a customer and the sale price \cite{Hwang2002}. They are
not only widespread mechanisms to sell goods, but have also been
shown applicable to task assignment, scheduling, or finding the
shortest path in a network with selfish nodes \cite{Brandt2005a}.
According to the goals and decision strategies, the electronic
auction protocols can be categorized into the sealed-bid auction
\cite{Lee2009,Franklin1996,Xiong2009}, the English auction
\cite{Chung2008,Hwang2002}, and the $(M+1)$st-price auction
\cite{Abe2003,Brandt2005a}. In an English auction, each bidder
offers the higher price one by one, and finally a bidder who offers
the highest price gets the desired goods. It is noted that all bid
values are published and any bidder easily knows the price position
of goods in English auction. Therefore, a bidder has the dominant
strategy for bidding, which places a little higher than a current
bid value. In this way, the competition principle well works and the
winning bid value reflects a market price. This is why an English
auction is the most familiar style of auctions. Therefore, this
study focuses on the English auction protocol, bringing up related
issues and methods.

Privacy is a crucial issue in designing the auction protocols. A
major reason why people may be hesitant to participate in auction
protocol themselves, is the worry that too much of their private
information is revealed. Furthermore, in the modern electronic
society, the information might get propagated to large numbers of
parties, stored in permanent databases, and automatically used in
undesirable ways. To solve this problem, we study the possibility of
designing the English auctions with communication round efficiency
in a way that preserves the bidders' privacy. Franklin and Reiter
\cite{Franklin1996} were among the first researchers to address
electronic auction with bid privacy. They covered many problems such
as secret sharing, digital cash and multicast as well as their own
primitive technique called verifiable signature sharing. Their
protocol successfully prevents a single auctioneer from altering a
bid or throwing an auction to a single bidder. Unfortunately, in
their protocol, all bids will be disclosed to all auctioneers after
the auction is closed. Kikuchi \textit{et al.} \cite{Kikuchi1999}
attempted to deal with such problems through secret sharing
techniques, but Sako \cite{Sako2000} pointed out that several
problems still remain in their work. Also, there are protocols where
bidders themselves jointly compute the auction outcome without
relying on trusted third parties at all
\cite{Brandt2005a,Brandt2005b}. The main advantage of these
protocols is that they are fully private, \textit{i.e.}, when
relying on computational intractability assumptions, no coalition of
parties is capable of breaching privacy. The drawbacks implied by
such a model are low robustness and relatively high computational
and communication complexity. Chang \textit{et al.} \cite{Chang2003}
and Jiang \textit{et al.} \cite{Jiang2005} proposed anonymous
electronic auction protocols based on the deniable authentication.
Nevertheless, in these protocols, the auctioneer must verify the
identity and bid price of all bidders one by one during the bidding
stage to ensure the legality of a bidder and the integrity of the
bid price. So these protocols will pose a heavy computation overhead
for the server at the auctioneer's end. Omote \textit{et al.}
\cite{Omote2001,Omote2002} initially proposed electronic English
auction which realize both anonymity of bidders and traceability by
employing bulletin boards. However, their method does not publicize
bidder information because publishing such information compromises
privacy, including anonymity, fairness and non-linkability among
various auction rounds, etc. Sakurai \textit{et al.}
\cite{Sakurai2000}, Nguyen \textit{et al.} \cite{Nguyen2000} and Lee
\textit{et al.} \cite{Lee2009} proposed anonymous and non-repudiate
auction protocols based on group signature respectively. Although
they realized the privacy-preserving auction protocol efficiently,
the auction manager have to maintain \emph{a black list}, which is
the list of participants that have their memberships revoked. Hence,
each bidder has to spend additional time on verifying whether the
other bidders had been revoked or not. Furthermore, when the number
of revoked members in the black list is larger than some threshold,
the protocol requires every remaining bidder to renew their secret
membership key and updating public information. To solve this
problem, Xiong \textit{et al.} \cite{Xiong2009} proposed an
anonymous auction protocol based on the ring signature, where the
bidder can be easily removed from the system. Whereas, taking round
efficiency into consideration, Xiong \textit{et al.}'s protocol is
much more costly than the previously protocols.

In this paper, we propose an English auction protocol with
privacy-preserving based on revocable ring signature
\cite{Shacham2007}. In addition to satisfy the above properties, our
protocol has the following unparalleled features: (a) The proposed
protocol can efficiently evict the malicious bidder instead of
maintaining the black list or updating the public information; (b)
Our protocol has low round communication complexity.

The remainder of this paper is organized as follows. The next
section presents background information related to English auction
protocol. Section \ref{sec3} details the proposed auction protocol,
followed by the security analysis and the performance analysis in
Section \ref{sec4}. Section \ref{sec5} concludes this paper.

\section{Preliminaries}
\label{sec2}
\subsection{Desired requirements}

Recently, the need for privacy has been a factor of increasing
importance in auction design and various schemes to ensure the safe
conduction of English auctions have been proposed
\cite{Brandt2005a,Brandt2005b,Chung2008,Franklin1996,Hwang2002,Juels2002,Kikuchi1999,Lee2009,Nguyen2000,Omote2001,Omote2002,Sako2000,Suzuki2000,Sakurai2000,Xiong2009}.
Meanwhile, any bid does not allow to be canceled in the case of
English auction. Because the highest bid may be insignificant if a
bid can be canceled in an English auction. Therefore, in an
electronic English auction, it is the most important to satisfy the
following two properties simultaneously: (a) Anonymity and (b)
Traceability. Although any bidder can participate anonymously, it is
necessary to identify a winner after the bidding phase without
winner's help. This means that every bid placed in an English
auction must be authorized while maintaining anonymity. In the
following, we summarize the requirements of electronic auction from
the researches of Chen\cite{Chen2004}, Chang and
Chang\cite{Chang2003}, and Chung \textit{et al.}\cite{Chung2008},
and Omote and Miyaji\cite{Omote2001,Omote2002}:

\begin{enumerate}
  \item Anonymity: Nobody including the authority itself can identify the losing
bidders even after the opening phase.
  \item Traceability: The cooperation of registration manager (AM) and auction manager
(RM) can identify the malicious bidder. In this way,
  the malicious bidder will be removed from the system. Note that an electronic auction has mainly two entities, the
RM who treats the registration of bidders, and the AM who holds
auctions.
  \item Unforgeability: Nobody can impersonate a certain bidder.
  \item Fairness: all bids should be fairly dealt with.
  \item Public verifiability: Anybody can publicly verify that a winning
bid is the highest value of all bids and publicly confirm whether
a winner is valid or not.
  \item Unlinkability among plural auctions: nobody
can link the same bidder¡¯s bids among plural auctions.
  \item Robustness: Even if a bidder sends an invalid bid, the auction
process is unaffected. \item One-time registration: any bidder can
participate in plural auctions by only one-time registration.
  \item Efficiency: The protocol should be efficient from the viewpoints
of computation and communication.
\end{enumerate}

\subsection{Bilinear Maps}
\label{bilinearmaps}

Since bilinear maps work of composite order as the basis of our
proposed scheme in this paper, we briefly introduce the bilinear
maps \cite{Boneh2005} in this section. Let $n$ be a composite with
factorization $n=pq$. We have

\begin{itemize}
    \item $\mathbb{G}$ is a multiplicative cyclic group of order $n$;
    \item $\mathbb{G}_{p}$ is its cyclic order-$p$ subgroup, and
    $\mathbb{G}_{q}$ is its cyclic order-$q$ subgroup;
    \item $g$ is a generator of $\mathbb{G}$, while $h$ is a
    generator of $\mathbb{G}_{q}$;
    \item $\mathbb{G}_T$ is a multiplicative group of order $n$;
    \item $\hat{e}:\mathbb{G}\times \mathbb{G}\rightarrow
\mathbb{G}_{T}$ is an efficiently computable map with the
following properties:
\begin{itemize}
    \item Bilinearity: For all $u,v\in \mathbb{G}$, and $a,b\in
\mathbb{Z}$, $\hat{e}(u^{a},v^{b})=\hat{e}(u,v)^{ab}$.\item
Non-degeneracy: $\hat{e}(g,g)= \mathbb{G}_{T}$ whenever
$<g>=\mathbb{G}$.
\end{itemize}
\item $\mathbb{G}_{T,p}$ and $\mathbb{G}_{T,q}$ are the
$\mathbb{G}_{T}$-subgroups of order $p$ and $q$, respectively;
\item the group operations on $\mathbb{G}$ and $\mathbb{G}_T$ can
be performed efficiently; and \item bitstrings corresponding to
elements of $\mathbb{G}$ and of $\mathbb{G}_T$ can be recognized
efficiently.
\end{itemize}

\subsection{Ring signature}

Ring signature, introduced by Rivest, Shamir and Tauman
\cite{Rivest2001}, is characterized by two main properties:
anonymity and spontaneity. Anonymity in ring signature means
$1$-out-of-$n$ signer verifiability, which enables the signer to
keep anonymous in these ``rings" of diverse signers. Spontaneity is
a property which makes distinction between ring signatures and group
signatures \cite{Chaum1991,Boneh2004}. Group signature allows the
anonymity of a real signer in a group to be revoked by a trusted
party called group manager. It also gives the group manager the
absolute power of controlling the formation of the group. Ring
signature, on the other hand, does not allow anyone to revoke the
signer anonymity, while allowing the real signer to form a group
(also known as a ring) arbitrarily without being controlled by any
other party. Since Rivest \textit{el al.}'s scheme, many ring
signature schemes have been proposed
\cite{Bresson2002,Abe2002,Wong2003,Boneh2003,Dodis2004}. Inspired by
the compact group signature\cite{Boyen2006}, Shacham and Waters
\cite{Shacham2007} proposed an efficient ring signature, which can
be proved secure in the standard model. Also inspired by the group
signature \cite{Boyen2006}, we remark that the anonymity in this
ring signature can be revoked by the trusted authority in the same
way like \cite{Boyen2006}. That is to say, the signature allows a
real signer to form a ring arbitrarily while allowing a trusted
authority to revoke the anonymity of the real signer. In other
words, the real signer will be responsible for what is has signed as
the anonymity is revocable by authorities while the real signer
still has the freedom on ring formation. In this paper, we propose a
conditional privacy-preservation English auction protocol with round
efficiency based on this modified ring signature scheme in
\cite{Shacham2007}.

\section{The proposed English auction protocol}
\label{sec3}

This section describes in detail our efficient privacy-preserving
auction protocol. In a high level description, the auction system
works as follows. To enrol in the system, each bidder contacts the
registration manager and registers his own public key and
corresponding real identity. (Through this way, the key escrow
problem will be solved. That is to say, the registration manager
can't frame an innocent bidder by forging the bidder's signature).
After confirming the validity of bidder's identity and public key,
registration manager will publish the public key of the bidder on
the Bulletin Board System (BBS). Each bidder collects the public
keys of other bidders from BBS managed by registration manager. Then
for each auction, a bidder can bid a value by generating a ring
signature on the bid on behalf of this set of public keys. (We
remark that the bidder's public key must be included in this set of
public keys). At the deadline, the identity of the bidder, who posts
the highest bid, is retrieved using the revocation procedure. Bidder
privacy is protected due to the anonymity and unlinkability
properties of the underlying ring signature scheme.

\begin{figure}[!b]
\centering
\includegraphics[scale=0.6]{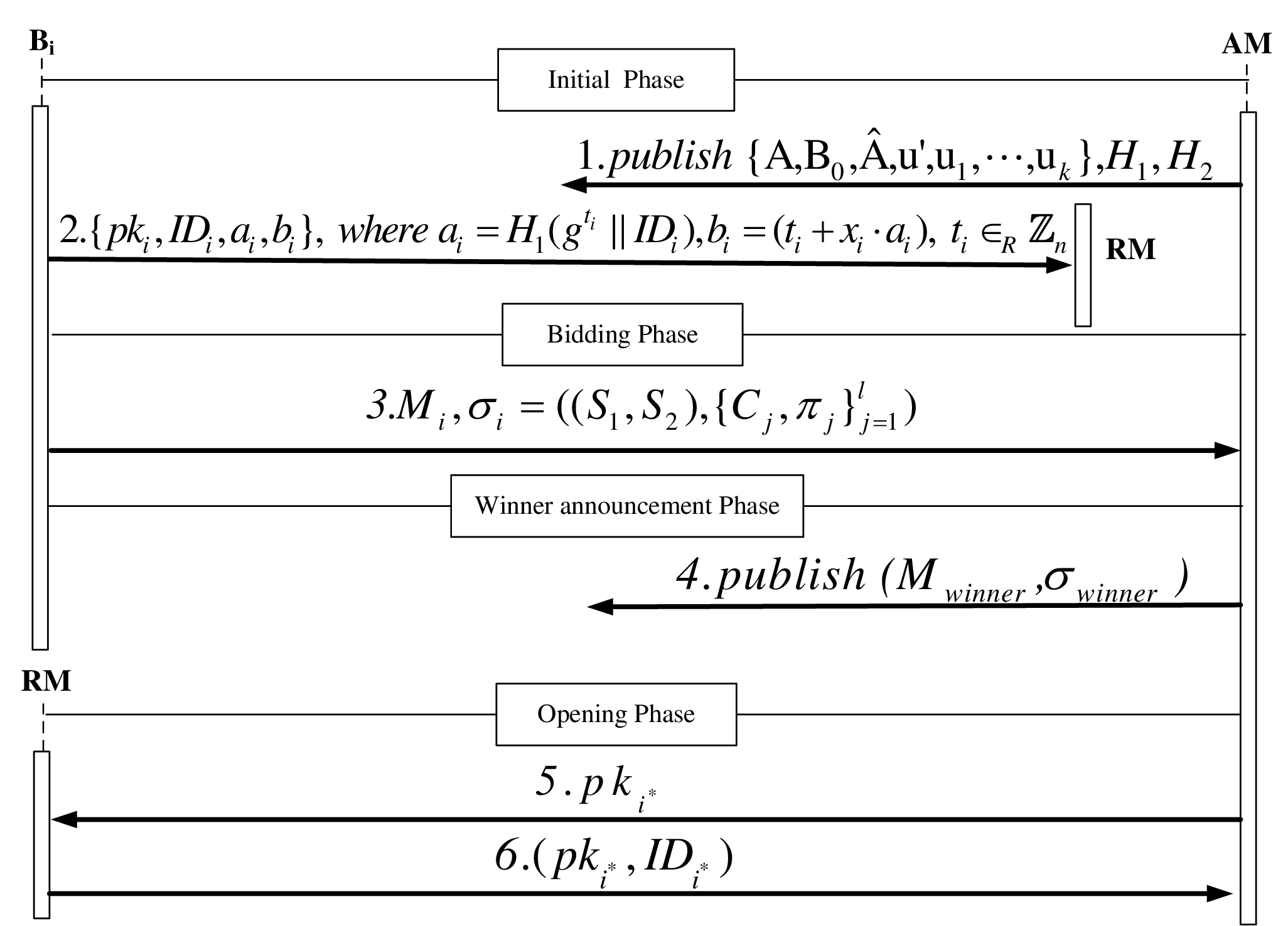}
\caption{English Auction procedure based on Ring signature}
\label{fig1}
\end{figure}

The proposed scheme includes the following four phases: Initial
phase, Bidding phase, Winner announcement phase, and Opening
protocol. The notations used throughout this paper are listed in
Table \ref{tbl1}. Let $B=\{B_1,\cdots,B_l\}$ be a set of $l$ bidders
who take part in an auction and offer a price. Let RM be
Registration Manager who manages the participants of auctions and AM
be Auction Manager who holds an auction and opens the real identity
of the bidder with RM. We assume that these two authorities RM and
AM do not collude together. Figure \ref{fig1} illustrates the
auction procedure.
\begin{table}[h]\caption{Notations}
\label{tbl1}
\begin{tabular}{ll}
Notations  &  Descriptions \\
\hline
RM:  &   Registration Manager who manages the participants of auctions by\\
     &    controlling the BBS. \\
AM:  &   Auction Manager who publishes $\{A,B_{0},\hat{A},u',u_{1},\cdots,u_{k}\}$ and keeps\\
     &   the tracing key $q\in \mathbb{Z}$ secret. \\
$B_{i}$:  & bidder who has its own secret key $pk_{i}=g^{x_{i}}$ and public key $sk_{i}=A^{x_{i}}$, \\
     &    where $x_{i}\in_R\mathbb{Z}_{n}$ \\
$ID_{i}:$ & The real identity of the bidder $B_{i}$\\
$S=\{pk_1,\cdots,pk_l\}$  &  $l$ public key of corresponding bidders\\
$M_i:$ & A bid generated by the bidder $B_{i}$\\
$\mathcal{H}_{1}(\cdot):$ & A hash function such as $\mathcal{H}_{1}:\{0,1\}^{*}\rightarrow \mathbb{Z}_{n}$\\
$\mathcal{H}_{2}(\cdot):$ & A hash function such as $\mathcal{H}_{2}:\{0,1\}^{*}\rightarrow \{0,1\}^{k}$\\
$a\parallel b$ & String concatenation of $a$ and $b$\\
\hline
\end{tabular}
\end{table}


\subsection{Initial phase}

Prior to the bidding phase, the AM sets up the system parameters
and publishes it as follows:

The AM first constructs a group $\mathbb{G}$ of composite order
$n=pq$ as described in section \ref{bilinearmaps} above. It then
chooses exponents $a,b_{0}\in_R\mathbb{Z}_{n}$ and sets $A=g^{a}$
and $B_{0}=g^{b_{0}}$ and $\hat{A}=h^{a}$. Let
$\mathcal{H}_{1}:\{0,1\}^{*}\rightarrow \mathbb{Z}_{n}$ and
$\mathcal{H}_{2}:\{0,1\}^{*}\rightarrow \{0,1\}^{k}$ be two
collision-resistant hash functions respectively. The AM picks hash
generators $u',u_{1},\cdots,u_{k}\in_R\mathbb{G}$. The published
parameters includes a description of the group $\mathbb{G}$ and of
the collision-resistant hash functions
$\mathcal{H}_{1},\mathcal{H}_{2}$, along with $(A,B_{0},\hat{A})$
and $(u',u_{1},\cdots,u_{k})$. Note that the auction manager AM's
tracing key is $q\in \mathbb{Z}$.

After receiving the public parameters from the AM, the RM is in
charge of checking the bidders' public key and identity as
follows:

\begin{itemize}
  \item The bidder $B_{i}$ first chooses $x_{i}\in_R\mathbb{Z}_{n}$; sets $pk_{i}=g^{x_{i}}\in\mathbb{G}$ as its private key, and $sk_{i}=A^{x_{i}}\in\mathbb{G}$ as its public
  key.
  \item $B_{i}$ randomly selects an integer
  $t_{i}\in_{R}\mathbb{Z}_{n}$ to determine the verification
  information of $pk_{i}$: $a_{i}=\mathcal{H}_{1}(g^{t_{i}}\parallel ID_{i})$ and $b_{i}=(t_{i}+x_{i}\cdot
  a_{i})$. Then $B_{i}$ transmits $\{pk_{i},ID_i,a_{i},b_{i}\}$ to RM
  over a secure channel.
  \item After receiving $\{pk_{i},ID_i,a_{i},b_{i}\}$, RM confirms the validity of the bidder's identity and checks whether the following equation holds:
  $$a_{i}\stackrel{?}{=}\mathcal{H}_{1}((g^{b_{i}}\cdot pk_{i}^{-a_{i}})\parallel ID_{i})$$
  If it holds, then $\{pk_{i},ID_i\}$ is identified as the valid public key and identity. Otherwise, it will be
  rejected. After that, RM keeps the relation $\{pk_{i},ID_i\}$ secret and publishes $pk_{i}$ on the BBS.
\end{itemize}

\subsection{Bidding phase}
\label{biddingphase}

%

In one round of auction, bidder $B_{i}$ signs his bid $M_i$ before
sending it out. Suppose $S=\{pk_{1},\cdots,pk_{l}\}$ is the set of
public keys and it defines the ring of public keys. We assume that
all public keys $pk_{i}$, $1\leq i \leq l$ and their corresponding
private keys $sk_{i}$'s are generated by the corresponding bidders,
and $i^{*}$ ($1\leq i^{*} \leq l$) is the index of the actual
bidder. The signature generation algorithm
$Sig(S,sk_{i^{*}},M_{i^{*}})$ is carried out as follows.

\begin{enumerate}
  \item Compute $\mathcal{H}_{2}(M_{i^{*}},S)=(m_{1},\cdots,m_{k})$. Define
$\{f_{i}\}_{i=1}^{l}$ as $$f_{i}=\left\{%
\begin{array}{ll}
    1, & \hbox{if $i=i^{*}$;} \\
    0, & \hbox{otherwise.} \\
\end{array}%
\right.    $$
  \item For each $i$, $1\leq i \leq l$, choose a random exponent
$e_{i}\in_R\mathbb{Z}_{n}$ and set
$C_{i}=(pk_{i}/B_{0})^{f_{i}}h^{e_{i}}$ and
$\pi_{i}=((pk_{i}/B_{0})^{2f_{i}-1}h^{e_{i}})^{e_{i}}$.
  \item Compute $C=\prod_{i=1}^{l}C_{i}$ and $e=\sum_{i=1}^{l}e_{i}$.
  \item Finally, choose $r\in_R\mathbb{Z}_{n}$ and compute
$S_{1}=sk_{i^{*}}\cdot (u'\prod_{j=1}^{k}u_{j}^{m_{j}})^{r}\cdot
\hat{A}^{e}$ and $S_{2}=g^{r}$.
\end{enumerate}

The signature $\sigma_i$ of $M_{i}$ with respect to $S$ is
$\{(S_{1},S_{2}),\{C_{j},\pi_{j}\}_{j=1}^{l}\}$. After generating
the ring signature on his bid successfully, $B_{i}$ will sends
$\{M_i,\sigma_i\}$ to the AM.



\subsection{Winner announcement phase}
\label{4-3}

After the deadline, AM chooses the highest (or the most suitable)
bid and runs the verify procedure. If the output is yes, AM accepts
the bid as the winning bid. Otherwise, AM repeats the process for
the remaining bids. The process of verifying the signature of bid is
as follows:

\begin{enumerate}
  \item Compute $\mathcal{H}_{2}(M,S)=(m_{1},\cdots,m_{k})$.
  \item For each $i$, $1\leq i \leq l$, check whether
  $\hat{e}(C_{i},C_{i}/(pk_{i}/B_{0}))\stackrel{?}{=}\hat{e}(h,\pi_{i})$
  holds or not. If any of the $C_{i}$ is invalid, reject.
  Otherwise, set $C=\prod_{i=1}^{l}C_{i}$. Accept if the following
  equation is satisfied:
  $\hat{e}(A,B_{0}C)\stackrel{?}{=}\hat{e}(S_{1},g)\hat{e}(S_{2}^{-1},u'\prod_{j=1}^{k}u_{j}^{m_{j}})$
\end{enumerate}

Once a winning bid is determined, AM posts the bid of the winning
bidder on the BBS along with the ring signature on this bid.

\subsection{Open protocol}
\label{secIVD}

If $B_i$ repudiates his bid or simply crashes, $AM$ invokes the open
protocol, which is two-party protocol between $AM$ and $RM$ for
opening the real identity of the bidder. At the beginning of this
protocol, AM checks the validity of the signature and then uses its
tracing key $q\in \mathbb{Z}$ and determines if
$$(C_{i})^{q}\cdot B_{0}\stackrel{?}{=}pk_{i}$$ for some $i$, $1\leq i\leq l$. If the equation holds at, say when $i=i^{*}$, then AM sends the
$pk_{i^{*}}$ to RM.

After receiving $pk_{i^{*}}$, RM looks up the record
$(pk_{i^{*}},ID_{i^{*}})$ to find the corresponding identity
$ID_{i^{*}}$ meaning that bidder with identity $ID_{i^{*}}$ is the
real bidder. The RM then evicts bidder $ID_{i^{*}}$ from auction and
$pk_{i^{*}}$ from BBS if this bidder is malicious. Otherwise, if
this bidder wins the auction, RM will send $(pk_{i^{*}},ID_{i^{*}})$
to the AM.

\section{Analysis}
\label{sec4}

The security and efficiency of auction protocol is analyzed in
this section. It will be shown that the protocol is fair, publicly
verifiable and achieves conditional privacy-preserving,
unlinkability, and robust.

\subsection{Security Analysis}

\noindent\textbf{Identity privacy preservation}: There is no single
authority who can break anonymity.

Given a valid ring signature $\sigma$ of some message, it is
computationally difficult to identify the actual bidder by any
participant in the system except the cooperation of RM and AM. If
there exists an algorithm which breaks the signer anonymity of the
construction in Section \ref{biddingphase}, then the Subgroup Hiding
(SGH) assumption would be contradicted \cite{Shacham2007}.
Furthermore, only RM knows the relation between the $pk_{i}$ and
bidder's real identity $ID_{i}$. So, only the cooperation of RM and
AM can break the bidder's anonymity.

\noindent\textbf{Non-repudiation}: No bidder can deny he had
submitted his bid.

Given the signature, the AM who knows the tracing key $q$,
  can trace the public key of a malicious bidder using the Dispute protocol described in section \ref{secIVD}. Besides, the
  tracing process carried by the AM does not require any
  interaction with the malicious bidder. With the cooperation of RM,
  the real identity of the malicious bidder can be revealed.

\noindent\textbf{Unforgeability}: In our protocol nobody can
impersonate any other bidder to make a bid.

According to \cite{Shacham2007}, the ring signature is unforgeable
with respect to the insider corruption if Computational
Diffie-Hellman (CDH) problem is hard. So, in our proposed scheme,
the bid along with the ring signature can only be generated by the
valid ring members.

\noindent\textbf{Fairness}: Our protocol has fairness of bidder
since the bidder is anonymous during the auction.

Fairness of bidder in an electronic auction means that any bid is
fairly accepted by AM. Generally, in an electronic English auction,
fairness of bidder depends on AM. Our protocol can avoid unfairness,
such as AM repudiates any bidding by a certain bidder, because the
bidding is done anonymously.

\noindent\textbf{Public verifiability}: It is public verifiable that
the price of the successful bid is higher than any other bids.

In our protocol, anyone can simulate the procedure to verify the
validity of bids using the information on the BBS. Since all the
information necessary to decide the auction result is published on
the BBS, anyone can verify the auction result.

\noindent\textbf{Unlinkability among plural auctions}: It is
impossible to link the same bidder among plural auctions.

Unlinkability is the basic property related to ring signature: two
ring signatures issued by the same signer are unlinkable in any way,
except the very fact that this signer appears in the set of public
keys of both ring signatures
\cite{Abe2002,Boneh2003,Bresson2002,Dodis2004,Shacham2007,Wong2003}.
So nobody can link two signatures among plural auctions.

\noindent\textbf{Robustness}: Malicious cheating and crashing can be
recovered.

Misbehavior takes place from time to time as a result of either
intentional malicious behaviors (e.g., attacks) or hardware
malfunctioning. It is less difficult to prevent misbehavior of
unauthorized users of auction protocols (i.e., outsiders) since
legitimate users can simply ignore the messages injected by
outsiders by means of authentication. This is one reason that we say
ring signature is the building block of auction protocols. On the
contrary, misbehavior of legitimate users of auction protocols
(i.e., insiders) is more difficult and complex to prevent, the
reason being that insiders possess the legitimate public/private key
pairs to perform authentication with peer bidders who can be easily
tricked into trusting the insiders. Consequently, the insiders'
misbehavior will have much larger impact on the network and be more
devastating. Fortunately, the opening phase can be employed to
detect such misbehavior and misbehaving users will be evicted
accordingly.

\noindent\textbf{One-time registration}: Any bidder can take part in
plural auctions as a valid bidder in one-time registration of public
key, maintaining anonymity for RM, AM, and other bidders.

Note that the honest bidder can get the public key of ring members
(a set of bidders) required to generate the ring signature
arbitrarily from the BBS without any interaction with any other
bidders, RM or AM in the system. So, the honest bidder can take
part in plural auctions in one-time registration.


\subsection{Efficiency Analysis}

In our protocol, the computational costs and communication overheads
on AM and RM are not stringent since these entities are
resource-abundant in nature. We are interested in the computational
costs and communication overheads at bidders which are least
powerful in our system. We use table \ref{efficiency} to show the
performance analysis of our protocol and \cite{Xiong2009}. For
convenience, we define the following notations: $T_e$ (the time for
one exponentiation computation); $T_m$ (the time for one modular
multiplication computation); $T_i$ (the time for one inverse
computation); $T_h$ (the time for executing the adopted one-way hash
function); $T_{Enc}$ (the time for executing the encryption
function); $T_{Dec}$ (the time for executing the decryption
function). The parameter $l$ and $k$ are used to denote the number
of public keys in the generation of ring signature and the length of
the output of hash function $\mathcal{H}_2$ respectively. It is
obvious that our proposed protocol possesses the advantages of
\cite{Xiong2009} due to the computation efficiencies and
communication rounds.



\begin{table}[h]
\caption{The comparison of efficiency}\label{efficiency}
  \begin{center}
  \begin{tabular}{|c|c|c|} \hline
     & Xiong et al.'s schem\cite{Xiong2009} &  Our scheme\\ \hline
    Initial Phase & {\small  $T_e+T_{Enc}+T_{Dec}$} &  {\small  $3T_e+T_h+T_m$}\\  \hline
    Pre-Bidding Phase & {\small  $(3l-2)T_e+lT_h+(2l+1)T_m$} &  \\ \hline
    Bidding Phase & {\small  $(3l-2)T_e+lT_h+(2l+1)T_m$} & {\small$T_h+(5l+k+2)T_e+(5l+k+1)T_m+2lT_i$} \\ \hline
    Winner announcement Phase & {\small  $(3l-2)T_e+lT_h+(2l+1)T_m$} &
    \\ \hline
   Rounds & {\small$4$} & {\small$2$}
    \\
    \hline
   \end{tabular}
  \end{center}
\end{table}


In addition, our protocol does not need to maintain \emph{a black
list} which is the list of evicted participants, different from the
auction protocol based on group signature
\cite{Sakurai2000,Nguyen2000,Lee2009}. Furthermore, our protocol
does not require every remaining bidder to renew their secret
membership key and updating public information. Thus, our protocol
is more practical than \cite{Sakurai2000,Nguyen2000,Lee2009}.

\section{Conclusions}
\label{sec5}

A privacy preserving English auction protocol with round efficiency
based on a modified ring signature has been proposed in this paper.
We demonstrate that proposed protocol does not only provide
conditional privacy, a critical requirement in English auction
protocols, but also able to improve efficiency in terms of
communication round complexity and identity tracking in case of a
dispute. Meanwhile, our proposed solution can achieve one-time
registration: that is to say, the bidder can take part in plural
auctions in one time registration.

\end{document}